\documentclass[11pt]{article}
\usepackage{textcomp}
\usepackage{pdfsync}
\usepackage{fancyhdr}
\usepackage{amssymb}
\usepackage{amsmath}

\usepackage{graphicx}
\usepackage{latexsym}
\usepackage{appendix}
\usepackage{srcltx}
\def\CD{{\cal D}}

\def\CH{{\cal H}}

\def\CO{{\cal O}}

\def\CT{{\cal T}}

\def \w {\omega}

\def \w {\omega}

\def\centeron#1#2{{\setbox0=\hbox{#1}\setbox1=\hbox{#2}\ifdim
   \wd1>\wd0\kern.48\wd1\kern-.48\wd0\fi
   \copy0\kern-.48\wd0\kern-.48\wd1\copy1\ifdim\wd0>\wd1
   \kern.48\wd0\kern-.48\wd1\fi}}

\def\JHEP{JHEP~}

\def\PRL{Phys. Rev. Lett.~}
\def\PR {Phys. Rev.~}
\def\CQG {Class. Quant. Grav.~}
\def\PL {Phys. Lett.~}

\newcommand{\beq}{\begin{equation}}
\newcommand{\eeq}{\end{equation}}
\newcommand{\bea}{\begin{eqnarray}}
\newcommand{\eea}{\end{eqnarray}}
\newcommand{\ba}{\begin{array}}
\newcommand{\ea}{\end{array}}

\newcommand{\nn}{\nonumber}

\newcommand{\half}{\frac{1}{2}}

\begin{document}

\hskip3cm

\hskip12cm{CQUeST-2013-0623}
\vskip3cm

\begin{center}
 \LARGE \bf    Some properties of the de Sitter black holes in three dimensional spacetime
\end{center}

\vskip2cm

\centerline{\Large Yongjoon Kwon\footnote{ykwon@sogang.ac.kr}\,,~~Soonkeon Nam\footnote{nam@khu.ac.kr}\,,
~~Jong-Dae Park\footnote{jdpark@khu.ac.kr} 
}
\hskip2cm

\begin{quote}
Department of Physics and Research Institute of Basic Science, Kyung
Hee University, Seoul 130-701, Korea$^{1,2,3}$

Center for Quantum Spacetime, Sogang University, Seoul 121-741,
Korea$^1$
\end{quote}

\hskip2cm

\vskip2cm

\centerline{\bf Abstract}  We investigate the physical properties of the de Sitter spacetime and new type-de Sitter black holes in new massive gravity, a higher derivative gravity theory in three dimensions. We calculate thermodynamic quantities and check that the first law of thermodynamics is satisfied. In particular, we obtain the energies of the de Sitter spacetime and new type-de Sitter black holes from the renormalized Brown-York boundary stress tensor on the Euclidean surfaces at late temporal infinity. We also obtain the quasinormal modes and by using them we find that the entropy spectra are equally spaced via semi-classical quantization.
  \thispagestyle{empty}
\renewcommand{\thefootnote}{\arabic{footnote}}
\setcounter{footnote}{0}
\newpage

%%%%%%%%%%%%%%%%%%%%%%%%%%%%%%%%%%%%%%%%%%%%%%%%%%%%%%%%%%%%%%%%%%%%%%%
\section{Introduction}
%%%%%%%%%%%%%%%%%%%%%%%%%%%%%%%%%%%%%%%%%%%%%%%%%%%%%%%%%%%%%%%%%%%%%%%

There have been many studies in three dimensional gravity theories over the past
few decades. It is well known that the gravity theory with the cosmological 
constant and gravitational 
Chern-Simons term (topologically massive gravity; TMG) has massive graviton 
modes and black hole solutions, such as BTZ and warped AdS black holes   
\cite{Deser:1982vy, Deser:1981wh, Hotta:2008, Vuorio:1985, Percacci:1987, Gurses:1994, Nutku:1993eb, Anninos:2008fx, Moussa:2008sj}. 
A few years ago, Bergshoeff, Hohm and Townsend have introduced a new gravity theory, 
so-called new massive gravity (NMG), 
that consists of the Einstein-Hilbert term, cosmological constant and 
higher curvature terms which have a specific combination of the square of 
the Ricci scalar and the square of the Ricci tensor 
\cite{Bergshoeff:2009hq, Bergshoeff:2009aq, Bergshoeff:2009tb, Bergshoeff:2009fj}. 
The one of the main 
purposes of the introduction of this NMG theory can be considered as the 
non-linear completion of the Pauli-Fierz massive graviton theory. 
Contrary to the TMG theory, it has even parity symmetry 
and two massive graviton modes. This theory with negative cosmological constant
also allows black hole solutions such as BTZ, warped AdS and new type black holes 
\cite{Clement:2009gq, Oliva:2009ip, Giribet:2009qz, Ghodsi:2010gk, Ahmedov:2010em}.
Another interesting side is to consider the realization of the $AdS/CFT$ 
correspondence \cite{Maldacena:1997re}. In the context of the consistency
with the holographic c-theorem, this theory may extend to the theory with the 
higher than square curvature terms \cite{Sinha:2010ai, Gullu:2010pc, Myers:2010tj}.

There have been lots of studies about NMG with negative cosmological constant, which has some asymptotically AdS solutions  \cite{Clement:2009gq,Oliva:2009ip,Giribet:2009qz,Sinha:2010ai,Gullu:2010pc,Hohm:2010jc,Giribet:2010ed,Grumiller:2009sn,
Alishahiha:2010bw,Nam:2010ma,Nam:2010dd,Kwon:2011ey}.
However, the investigation for NMG with positive cosmological constant has hardly done so far.
In Einstein-Hilbert gravity 
theory with positive cosmological constant, there exist asymptotically de Sitter black holes  with two horizons in $D \ge 4$.
One is the cosmological horizon and the other is the black hole horizon. These horizons can be 
represented by physical parameters, mass and cosmological constant.
In three dimensions the Schwarzschild-de Sitter(dS) spacetime has only one horizon and a conical singularity at the origin.
Even though it does not have a black hole horizon, we will keep the name since its metric corresponds to the three dimensional case of the Schwarzschild-dS metric in general dimensions.
Considering NMG theory, the new type-de Sitter(dS) black hole solution can be allowed and also
have two parameters even in the static situation likewise the new type black 
hole in asymptotically AdS case. Due to these two parameters, this black hole has both 
cosmological and black hole horizons. 

The main purpose of this paper is to study about properties of these asymptotically dS spaces 
in three dimensions. We investigate thermal properties, stability and entropy 
spectrum as a quantum property of these solutions. Firstly, we calculate  
entropies and energies of the Schwarzschild-dS spacetime and new type-dS black holes. With these results, the first law  of  thermodynamics is checked. Secondly, we calculate quasinormal modes(QNMs) of 
these spaces. The QNMs are the one of the useful tools for understanding about black
hole physics \cite{Nollert:1999ji, Kokkotas:1999bd}. 
Even though QNMs represent just classical damping
perturbation which propagates on some backgrounds, these can give us some information 
about quantum properties. 
By using quasinormal modes, for example, the quantization of the black hole can be obtained via semi-classical method. 
Moreover, it has been shown that the damping time scales of the black hole
on the AdS space correspond to the thermalization time scales of the conformal field 
theory on the AdS boundary \cite{KalyanaRama:1999zj, Horowitz:1999jd}.

This paper is organized as follows. In section 2, we briefly review about NMG theory 
and simply describe the Schwarzschild-dS and new type-dS black hole solutions. We also 
calculate thermal quantities such as Hawking temperatures, entropies and energies of these 
asymptotically dS spaces. In section 3, by solving the massive scalar equation on 
two asymptotically dS solutions, we obtain the exact QNMs of the Schwarzschild-dS, and 
the asymptotic QNMs of the new type-dS black holes by using monodromy method.
In section 4, we find the quantized 
entropy spectra of the two asymptotically dS solutions from the asymptotic QNMs.
  In section 5, we make conclusions and some 
comments. 

%%%%%%%%%%%%%%%%%%%%%%%%%%%%%%%%%%%%%%%%%%%%%%%%%%%%%%%%%%%%%%%%%%%%%%%

%%%%%%%%%%%%%%%%%%%%%%%%%%%%%%%%%%%%%%%%%%%%%%%%%%%%%%%%%%%%%%%%%%%%%%%
\section{The new massive gravity with positive cosmological constant}
%%%%%%%%%%%%%%%%%%%%%%%%%%%%%%%%%%%%%%%%%%%%%%%%%%%%%%%%%%%%%%%%%%%%%%

 In this section,   we will understand some thermodynamic properties of solutions in NMG with positive cosmological constant.  
For positive cosmological constant, the NMG action is given by 
\begin{equation} 
\label{NMGaction}
{\cal S} ={1 \over {16 \pi G }}\int d^3x\sqrt{-g}\bigg[ R -{2 \over \ell^2} + \frac{1}{m^2}K  \bigg]\,,
\end{equation}
where
$K$ is defined by
\beq
 K = R_{\mu\nu}R^{\mu\nu} -\frac{3}{8}R^2\,.
\eeq
Then, via the metric variation, the EOM is obtained as
\begin{equation}
   G_{\mu\nu} + {1 \over \ell^2} g_{\mu\nu} + \frac{1}{2m^2}K_{\mu\nu} 
        =0\,,
\end{equation}
 where
\begin{equation}
 K_{\mu\nu} = g_{\mu\nu}\Big(3R_{\alpha\beta}R^{\alpha\beta}-\frac{13}{8}R^2\Big)
                + \frac{9}{2}RR_{\mu\nu} -8R_{\mu\alpha}R^{\alpha}_{\nu}
                + \half\Big(4\nabla^2R_{\mu\nu}-\nabla_{\mu}\nabla_{\nu}R
                -g_{\mu\nu}\nabla^2R\Big) \nn\,.
\end{equation}
It can be easily checked that this has the Schwarzschild-dS spacetime which has only the cosmological horizon. In addition, it also has black hole solutions with multiple horizons like the new type black holes in AdS case of NMG.  In three dimensions, it is known that the Einstein gravity has no black hole solution with asymptotically dS space. However, we will notice that in NMG with positive cosmological constant such black holes, so-called new type-dS black holes, can exist at the special point. Throughout this paper, the unit of $c=\hbar=1$ will be used.

%%%%%%%%%%%%%%%%%%%%%%%%%%%%%%%%%%%%%%%%%%%%%
\subsection{Thermodynamics of the Schwarzschild-dS spacetime }
%%%%%%%%%%%%%%%%%%%%%%%%%%%%%%%%%%%%%%%%%%%%%%

First, we begin by considering the Schwarzschild-dS metric given by
\beq
\label{sdsmet}
ds^2= -\bigg(8 G M- {r^2 \over L^2}\bigg) dt^2+ {\bigg(8 G M- {r^2 \over L^2}\bigg)^{-1} } dr^2+r^2 d\phi^2 \,,
\eeq
where $L$ is the dS radius. 
To be a solution of the EOM, it is found that the following relation should be satisfied: 
\bea \label{schrel}
{1 \over \ell^2}= {1 \over L^2} \bigg(1+ {1 \over {4 m^2 L^2}} \bigg) \,.
\eea
The metric (\ref{sdsmet}) has the cosmological horizon at $ r_c = L \sqrt{8 G M} $ and the scalar curvature of this metric is given by
\bea R = \frac{6}{L^2} \,. 
\eea
The Hawking temperature at the cosmological horizon is obtained as
\bea
 T_{H}= \Big|{ \kappa_c \over {2 \pi}} \Big| ={ \sqrt{8 G M} \over {2 \pi L}}  = {r_c \over {2 \pi L^2}}  \,,
\eea
where $\kappa_c$ is the surface gravity at the cosmological horizon.

In the asymptotically dS case, there is another definition of temperature, called Bousso-Hawking(BH) temperature. This BH temperature is defined at the position $r_g$ where the gravitational acceleration vanishes  \cite{Bousso:1996au} as follows: 
\beq \label{bht}
T_{BH}=\Big| {{{\tilde \kappa}_c} \over {2 \pi}} \Big|  = {{f'(r_c)} \over {4 \pi \sqrt{f(r_g)}}}\, ,
\eeq
 where $f(r)=(8 G M- {r^2 / L^2})$, the surface gravity is defined in terms of a timelike killing vector with a normalization constant, $\xi=\gamma {{\partial } \over {\partial t}}$, as 
\bea
{\tilde \kappa}_c = {\sqrt {\xi^{\mu} \nabla_{\mu} \xi_{\nu} \xi^{\alpha} \nabla_{\alpha} \xi^{\nu} \over {-\xi^2}} }\Bigg|_{r=r_c}
\eea
and the normalization constant is given by $\gamma= 1/ {\sqrt{f(r_g)}}$ to satisfy $\xi^{\mu} \xi_{\mu}=-1$ at the reference point $r_g$. In this Schwarzschild-dS spacetime, the reference point is located at the origin $r=0$. 
Therefore, the BH temperature is calculated as
\bea
T_{BH}= {1 \over {2 \pi L}} \,.
\eea

The entropy of this spacetime can be obtained from the Wald entropy. For the black holes in higher derivative gravity, it is different from the area law in general. 
In NMG, the entropy of the Schwarzschild-dS is given by
\beq
\label{sdsen}
S_{W} = \frac{2 \pi r}{4G} \bigg[1 + \frac{1}{4m^2}\Big(R^{t}_{~t} + R^{r}_{~r} -3R^{\phi}_{~\phi}\Big)\bigg]_{r=r_c} = \Big(1 - \frac{1}{2m^2 L^2} \Big) \frac{\pi r_c  }{2G} \,.
\eeq

Now, let us find the energy of the Schwarzschild-dS in NMG. Indeed, there are some difficulties in defining the conserved charges in asymptotically dS space, since there are no spatial infinity and no asymptotic killing vector which is globally timelike.
In this paper, in order to calculate the conserved quantity, we will follow the method proposed in the work of   \cite{Balasubramanian:2001nb}. 
In AdS space, by calculating the Brown-York(BY) boundary stress tensor \cite{Brown:1992br} the energy of asymptotically AdS spacetime can be obtained. Through analogous procedure, the energy of asymptotically dS spacetime can be obtained by the Euclidean boundary stress tensor. The metric can be written in ADM form as follows:
\bea
ds^2=g_{\mu \nu}dx^{\mu} dx^{\nu}=-N_t^2 dt^2+h_{i j}(dx^{i}+V^{i}dt)(dx^{j}+V^{j}dt) \,,
\eea
where $h_{i j}$ is the induced metric on the surfaces of the fixed time. 
The spacetime boundaries, $\cal I^{\pm}$, are Euclidean surfaces at early and late time infinity.
Then the stress tensor can be obtained by variations of the action either on a late or early time boundary;
\bea
T^{+ i j} = {2 \over {\sqrt{h}}} {{\delta \cal S} \over {\delta h_{i j} }}, \qquad  T^{- i j} = {2 \over {\sqrt{h}}} {{\delta \cal S} \over {\delta h_{i j} }} \,.
\eea
In this paper, we will work on the Euclidean surfaces at $\cal I^+$. 
However, it should be noted that since for asymptotically dS space $r$ becomes timelike as $r \rightarrow \infty$, the calculation of the BY stress tensor will be very similar to the case of asymptotically AdS. To obtain the BY stress tensor, the generalized Gibbons-Hawking term is needed since we are dealing with not the Einstein gravity but the higher derivative gravity, NMG. 
To obtain the generalized Gibbons-Hawking(GH) term, the auxiliary field approach\footnote{One may note  that our convention is slightly different from~\cite{Hohm:2010jc}.} by Hohm and Tonni \cite{Hohm:2010jc} is used. The NMG action (\ref{NMGaction}) can be rewritten in terms of auxiliary field $f_{\mu\nu}$ as
\beq {\cal S} =\frac{1}{16 \pi G}\int d^3x\sqrt{-g}\bigg[  R -
\frac{2}{\ell^2} + f^{\mu\nu}G_{\mu\nu}-\frac{m^2}{4}\Big(f^{\mu\nu}f_{\mu\nu} - f^2\Big)  \bigg]\,.  \label{Action} \eeq
Then, the EOM is given by
\beq    G_{\mu\nu} +\frac{1}{\ell^2}g_{\mu\nu}  =   \CT^{B}_{\mu\nu}\,, \qquad   f_{\mu\nu} =\frac{2}{m^2}\Big(R_{\mu\nu} - \frac{1}{4}R g_{\mu\nu}\Big)\,,
        \eeq
where
\bea \!
     \CT^{B}_{\mu\nu} &=& \frac{m^2}{2}\Big[ f_{\mu\alpha} f^{\alpha}_{\nu} -f f_{\mu\nu} -\frac{1}{4}\Big(f^{\alpha\beta}f_{\alpha\beta} -f^2\Big)  g_{\mu\nu}\Big]  + \half f R_{\mu\nu} - \half Rf_{\mu\nu} -2 f_{\alpha (\mu}G^{\alpha}_{\nu)} + \half f^{\alpha\beta}G_{\alpha\beta} g_{\mu\nu} \nn \\
      && -\half \Big[ \CD^2f_{\mu\nu} + \CD_{\mu}\CD_{\nu} f -2\CD^{\alpha}\CD_{(\mu}f_{\nu)\alpha} + \Big(\CD_{\alpha}\CD_{\beta}f^{\alpha\beta} - \CD^2f\Big)g_{\mu\nu} \Big]\,.  \nn \eea
The auxiliary fields $f^{\mu\nu}$ can be decomposed as
\bea\label{auxliary} f^{\mu\nu} = \left(\ba{cc} s & h^j \\ h^i
& f^{ij} \ea\right) \,.  \nn \eea
With this decomposition the generalized GH term was obtained   in the form of
\beq
 {\cal S}_{GH} = \frac{\eta}{16 \pi G}\int d^2x \sqrt{-\gamma}\Big[2  K + \hat{f}^{ij}K_{ij} - \hat{f}K\Big]\,,
\eeq
where  $\eta$ is taken as $-1$ for the asymptotically dS case, while taken as $+1$ for the asymptotically AdS case, the definitions of $\hat{f}^{ij}$, $\hat{f}$ and covariant derivatives of them are given in  \cite{Hohm:2010jc}. Note that  the first term  is the GH term in pure Einstein gravity case.

In the AdS case of NMG, considering the generalized Gibbons-Hawking term, the BY boundary stress tensor was obtained in \cite{Hohm:2010jc}. For the asymptotically dS case, however, it should be slightly changed.
With the ADM-like split of the metric,
\bea
ds^2=\eta N^2 dr^2+\gamma_{ij}(dx^i +N^idr)(dx^j+N^j dr) \,,
\eea
 it is found that  the BY boundary stress tensor for the asymptotically dS case is given by
\bea
\label{BYT}
8 \pi G T^{i j}_{BY} &=& \Big(\eta  + \half \hat{s} -\eta \half \hat{f}\Big)(K\gamma^{ij} - K^{ij})  - \nabla^{(i} \hat{h}^{j)}+ \eta \half D_{r} \hat{f}^{ij}+ \eta K^{(i}_{k} \hat{f}^{j) k} \nn \\ 
&&+\gamma^{ij} \Big(\nabla_{k}\hat{h}^{k} -\eta \half D_{r}\hat{f}\Big) \,, 
\eea
where $i, j = t,\phi$.  
The extrinsic curvature is defined in terms of the normal vectors $K_{\mu \nu} =  \nabla_{(\nu} n_{\mu)}$,
where $\mu$ and $\nu$ are three dimensional indices.
This BY stress tensor (\ref{BYT}) is divergent as $r \rightarrow \infty$ as usual. In order to obtain the renormalized BY stress tensor,  we will consider the counter term in the form of 
\bea
 \label{counter}
{\cal S}_{ct}= {1 \over  {8 \pi G}} \int d^2 x \sqrt{-\gamma} \left(A+B \hat f+C\hat f^2+D f_{i j} f^{i j} \right)\,,
\eea
which leads to
\bea
8 \pi G T^{i j}_{ct}=  \left(A+B \hat f+C\hat f^2+D f_{i j} f^{i j} \right) \gamma^{i j}\,.
\eea
Therefore, the renormalized BY boundary stress tensor is given by
\bea
\label{RBYT}
8 \pi G \tilde T^{i j}  \equiv 8 \pi G(T^{i j}_{BY}+T^{i j}_{ct})|_{r \rightarrow \infty} \,.
\eea
From this renormalized stress tensor, the energy of the Schwarzschild-dS is obtained  with the metric expanded as 
\bea
ds^2=-{L^2 \over r^2}dr^2+{r^2 \over L^2}(dt^2+dx^2)+ \epsilon \bigg({{8 G M L^4} \over r^4} dr^2 - 8 G M dt^2 \bigg) +\CO(\epsilon^2) \nn \,,
\eea
where $x= L \phi$ and $\phi \in [0,2 \pi]$.
When the coefficients of the counter terms have the relation,
\bea
A-\frac{1}{L}+{1 \over {2 m^2 L^2 }} \left(4
   B+\frac{1}{L} +\frac{8 C+4 D}{m^2 L^2 } \right)=0 \,,
\eea
the divergent leading term of the BY tensor (\ref{BYT}) vanishes and 
the energy is obtained as
\beq 
\label{dsen}
E=  \int dx \tilde T_{t t}  =\left(1-\frac{1}{2 m^2 L^2 }\right) M \equiv \alpha M \,.
\eeq
As $m \rightarrow \infty$, the energy of the Schwarzschild-dS reduces to the Einstein case in  \cite{Balasubramanian:2001nb}. The overall factor is attributed to the effect of the higher derivative terms in NMG. 
When $M= 1/8G$, the metric (\ref{sdsmet}) becomes the pure dS spacetime of which the energy is $E_{dS}=\alpha /{8 G}$. This energy is the upper bound for the Schwarzschild-dS spacetime, since the existence of the conical singularity (or black hole in higher dimensional case) on the dS background reduces the energy of the pure dS space \cite{Balasubramanian:2001nb}. In other words, $E=E_{dS}-E_{M}$, where $E_M$ is the contribution of energy reduction from the occurrence of the conical singularity (or the existence of a black hole in higher dimensional case). This is related to the conjecture that the energy of any asymptotically dS space does not exceed the one of the pure dS space \cite{Balasubramanian:2001nb,Bousso:2000nf}. 

With the above calculated thermodynamic quantities, it can be easily checked that the first law of thermodynamics is satisfied.
\bea
T dS = dE \,.
\eea
%

%%%%%%%%%%%%%%%%%%%%%%%%%%%%%%%%%%%%%%%%%%%%%%%%%%%%
\subsection{Thermodynamics of the  new type-dS black holes}
%%%%%%%%%%%%%%%%%%%%%%%%%%%%%%%%%%%%%%%%%%%%%%%%%%%%
In this section, we will obtain the thermodynamic quantities for the new type-dS black holes by the same procedure used in the previous section.
As mentioned before, at the special point of  $m^2=-1/(2 L^2)=-1/\ell^2$, black hole solutions with asymptotically dS can exist in the NMG with positive cosmological constant. 
The metric of the new type-dS black holes is given by
\beq \label{nmg-metric}
  ds^2 =  -\left(-{r^2 \over  L^2} + b {r \over L} - c\right)dt^2 + {\left(-{r^2 \over  L^2} + b {r \over L} - c\right)^{-1}} {dr^2} + r^2 d\phi^2\,,
\eeq
where $L$ is dS radius. Note that the black hole parameters, $b$ and $c$, are dimensionless, and the  coordinates, $t$ and $r$, have length dimension.  
This corresponds to weaker boundary conditions than the dS counterpart of the Brown-Henneaux boundary conditions \cite{Brown:1986nw}.  Some different boundary conditions from the Brown-Henneaux ones have been also considered in three dimensional conformal gravity \cite{Afshar:2011qw}.
The scalar curvature of this metric is given by
\bea
\label{curvature}
 R = \frac{6}{L^2} - \frac{2b}{Lr}\,,
 \eea
which shows us that there is a curvature singularity at $r=0$ when $b\neq 0$.
The black hole solutions can be classified according to the ratio of the two horizons as follows: 
\begin{center}
\begin{tabular}  [hb] { | c | c | c | }
\hline
   $q \equiv r_h/  r_c $   &  $b$  and $c$    & $r_c$ and $r_h$   \\
 \hline
 $q=1$  & $ b>0,~  c>0$   ($b^2 =4c$ ) &   $ {r_c}  = { r_h} >0  $   \\
\hline
   $0<q<1$  & $b>0,~ c>0$  ($b^2 >4c$ )   & $ {r_c} > { r_h} > 0$  \\
\hline
   $q=0$  &  $b>0,~ c= 0$   &$ r_c > r_h=0$   \\
 \hline
   $-1<q<0$  & $b>0,~ c<0 $   & $r_c >0,~ r_h<0$  $(  {r_c}  >| { r_h} | )$ \\
 \hline
   $q = -1$  & $b=0,~ c<0$   & $r_c >0, ~r_h<0$  $(  {r_c} =- { r_h} )$ \\
\hline
   $q< -1$ & $b<0,~ c<0$  &   $r_c >0, ~r_h<0$  $(  {r_c}  < | { r_h} | )$ \\
\hline
\end{tabular}
\end{center}
As shown in the above table, when $0 <q <1$, the black hole has the black hole horizon as well as the cosmological horizon at $r_h= {L \over 2} (b - \sqrt{b^2-4c})$ and $r_c= {L \over 2} (b  + \sqrt{b^2-4c})$. When $q=-1$, the metric reduces to the Schwarzschild-dS spacetime, eq.(\ref{sdsmet}) with $c= -8GM$. 
However, note that when $q \le 0 $ except for $q=-1$, the scalar curvature (\ref{curvature}) is divergent at $r=0$, which implies the singularity is naked. Therefore this case is not considered as physically meaningful  one. 
This is different from the new type black holes in the NMG with negative cosmological constant \cite{Kwon:2011ey}, where the singularity at the origin is always hidden behind the event horizon.
Therefore, in the following we will only focus on the case of $0<q<1$ corresponding to the new type-dS black holes, since the case of $q=-1$ corresponding to the Schwarzschild-dS at the special point is already included in the previous section. 

The Hawking temperatures of this black hole at the cosmological and black hole horizons are given by
\bea
T^{h,c}_H= \Big| {{ \kappa_{h,c}} \over {2 \pi}} \Big|  
 =   \frac{1}{4 \pi L}\sqrt{b^2-4c}   =   \frac{r_c - r_h}{4 \pi L^2}\,.
\eea
Note that $\kappa_c <0$ at the cosmological horizon and $\kappa_h >0$ at the black hole horizon.
The Bousso-Hawking temperatures are calculated from the eq.(\ref{bht}) as
\bea
T^{h,c}_{BH} = {1 \over {2 \pi L}}\,.
\eea

From the Wald entropy, the entropy is obtained as
\bea
\label{ntentro}
 S^{h,c}_{W}  = \frac{2 \pi r}{4G} \Bigg[ 1+{1 \over {4 m^2}}\left( {{-2 r+2b L} \over {L^2 r}}\right) \Bigg] _{r=r_{h,c}}=\pm \frac{\pi L}{2G} \sqrt{b^2-4 c}  =\pm \frac{\pi }{2G}(r_c - r_h)\,,
\eea
where $+$ is for the cosmological horizon, $-$ is for the black hole horizon, and  $2m^2 L^2=-1$.
It seems strange since the entropy at the black hole horizon is negative. 
It can be inferred that this bizarre phenomenon comes from the higher derivative effect. 
But, it may be natural in physical sense that when the horizon size increases (or decreases), the entropy, which implies the information on the horizon, should increase (or decrease). For examples, let us fix the one of the  black hole parameters, $b$ and $c$, and vary the other one. Then if the black hole horizon, $r_h$, increases (or decreases), the entropy at the black hole horizon, eq.(\ref{ntentro}), also increases (or decreases). Therefore in this sense, the negative sign in the entropy seems plausible, even though there are some unallowed regime of the variations when both parameters are varied at the same time. 
Indeed, negative entropy phenomena have been observed in other higher derivative gravity theories \cite{Nojiri:2001pm,Cvetic:2001bk,Cai:2009ac}. 
However, it is still not clear what the physical meaning of the negative entropy is. 
Now, let us calculate the energy of the new type-dS black holes. 
The metric of the new type-dS black holes can be expanded as
\bea
ds^2= -{L^2 \over r^2}dr^2+{r^2 \over L^2}(dt^2+dx^2)- \epsilon \left( {{b L^3}  \over r^3} dr^2 + {{b r} \over L} dt^2 \right) +\epsilon^2 \left( {{c L^4} \over r^4} dr^2 + c dt^2 \right) +\CO(\epsilon^3) \,, \nn 
\eea
where $x= L \phi$ and $\phi \in [0,2 \pi]$.
With the same form of the counter term, eq.(\ref{counter}), the two divergent terms of the BY stress tensor (\ref{BYT}) vanish when the coefficients of the counter term are satisfied with the following relations:
\bea
&&A-\frac{1}{L}+{1 \over {2 m^2 L^2 }} \left(4
   B+\frac{1}{L} +\frac{8 C+4 D}{m^2 L^2 } \right)=0 \,, \nn \\
  &&  A-{3 \over {2 L}}+{1 \over {2 m^2 L^2}} \left( 6 B+{ 3 \over {2 L}}+{{16C+8 D} \over {m^2 L^2}} \right)=0\,,
\eea
where $2 m^2 L^2=-1$. 
Note that the second condition appears only when $b \neq 0$. 
Then, the energy of the new type-dS black holes is obtained as
\beq
\label{nten}
E=  \int dx \tilde T_{t t}  ={1 \over {16 G}}(b^2-4c) ={1 \over {16 G L^2}}(r_c - r_h)^2 \,.
\eeq
It should be mentioned that in order to eliminate an arbitrariness of the calculated energy  the condition, $C+D=0$, should be imposed. Otherwise, the energy depends on the coefficient $C$ and $D$ of the counter term.
Interestingly, contrary to the Schwarzschild-dS case, the new type-dS black holes do not have the upper bound in energy, since in the regime of the black hole parameters, $b^2 > 4c > 0$, the pure dS space is not allowed. 

It can be also checked that the above thermodynamic quantities are satisfied with the first law of thermodynamics,
\bea
T^{c} dS^{c} = dE \,.    
\eea

%%%%%%%%%%%%%%%%%%%%%%%%%%%%%%%%%%%%%%%%%%%%%%%%%%%%%%%%%%%%%%%%%%%%%%%%%%
\section{Quasinormal modes of asymptotically dS space}
%%%%%%%%%%%%%%%%%%%%%%%%%%%%%%%%%%%%%%%%%%%%%%%%%%%%%%%%%%%%%%%%%%%%%%%%%%
In this section, we will find the QNMs of the Schwarzschild-dS and the new type-dS black holes. The QNMs for a scalar field are obtained from the Klein-Gordon field equation by imposing an appropriate boundary condition.
By the separation of variables, $\Psi = R(r) e^{i \mu \phi} e^{i \w t}$,
the radial equation can be obtained from the EOM for a massive scalar field, given by 
\begin{equation}\label{wave-eq}
\nabla^2 \Psi - m^2 \Psi = 0\, ,
\end{equation}
where $m^2$ is the mass of the scalar field.
%

%%%%%%%%%%%%%%%%%%%%%%%%%%%%%%%%%%%%%%%
\subsection{Exact quasinormal modes }
%%%%%%%%%%%%%%%%%%%%%%%%%%%%%%%%%%%%%%
%
First, we will consider the Schwarzschild-dS spacetime of which metric is given by the eq.(\ref{sdsmet}).
Through the change of variable,
\bea
z={{r^2} \over {8GM L^2}}\,,
\eea
the radial equation can be rewritten as
\bea
\label{pseq}
R''(z) +\left(\frac{1}{z}+\frac{1}{z-1}\right)
   R'(z)+{1 \over {z (z-1)}} \left(\frac{\mu^2}{\chi z}+\frac{
   \w^2 L^2}{\chi (z-1)} +\frac{m^2 L^2}{4} \right) R(z) =0 \,, \nn 
\eea
where $\chi \equiv 32GM $ and $'$ denotes the
derivative with respect to  the variable $z$.
Defining $ R(z) \equiv z^{\alpha} (1-z)^{\beta} H(z)$, and choosing one of the values of $\alpha$ and $\beta$ in 
\bea 
\{\alpha, \beta \} = \bigg\{ \pm {\mu \over {\sqrt{\chi}} },~ \pm i {{\omega L} \over {\sqrt{\chi}}}  \bigg\} \,, \nn
\eea
the above differential equation becomes the hypergeometric differential equation as follows:
\bea
H''(z)&+&\left(\frac{2
   \beta+1}{z-1}+\frac{2 \alpha+1}{z}\right) H'(z) \nn \\
   &+& {1 \over {z (z-1) }} {\left( (\alpha+\beta)^2+(\alpha+\beta) +\frac{  m^2 L^2}{4}\right) H(z)}=0 \,.
\eea
Without loss of generality, we take the values of $\alpha=-{\mu / {\sqrt{\chi}}}$ and $ \beta=-i {{\omega L} / {\sqrt{\chi}}}$. 
At $z=0$, the Dirichlet boundary condition should be imposed since the effective potential in a Schr$\ddot {\rm o}$dinger-like wave equation given in the tortoise coordinates is divergent there.  
From the connection formula between local solutions near singularities, it is found that the coefficient given by gamma functions in the following relevant part of the solution near $z=1$ should vanish by the boundary condition that only  the outgoing modes  exist  at the cosmological horizon ($z=1$):
\begin{eqnarray}
R(z) \sim z^{-\alpha}  (1-z)^\beta {{\Gamma(c) \Gamma(c-a-b)} \over  {\Gamma(c-a)\Gamma(c-b)}} {}_2 F_1(a,b,a+b-c+1|1-z) \,,
\end{eqnarray}
where
\bea
a \equiv \beta-\alpha+\frac{1}{2}\left(1- \sqrt{1- m^2 L^2}\right)\!, ~ b  \equiv  \beta-\alpha+\frac{1}{2}\left(1+ \sqrt{1-  m^2 L^2} \right)\!, ~ c  \equiv   1-2\alpha \,. \nn
\eea
Therefore, the exact QNMs of the Schwarzschild-dS are obtained as
\bea \label{sdsqnm}
 \w= i {{\mu +\sqrt{8GM}(2n+1 \pm \sqrt{1-m^2 L^2}) } \over L} \,.
\eea
When $8GM=1$, it represents the QNMs of the pure dS spacetime.

Next, let us consider the QNMs of the new type-dS black holes of which the metric is given by the eq.(\ref{nmg-metric}). 
Through the change of variable,
\beq\label{chvar} z \equiv \frac{r-{r_h}}{{r_c}-{r_h}}\,, \eeq
the radial equation becomes
\begin{eqnarray}\label{eqradial}
&&  R''(z) + \bigg[\frac{1}{z}+\frac{1}{z-1}
+\frac{1}{z-z_0}
   \bigg] R'(z)  \nonumber \\
&&    + {{R(z)} \over {z (z-1)(z-z_0)}}
   {\bigg[  {{\w^2 L^4 z_0} \over { {(r_c -r_h)^2} z} }+{{\w^2 L^4 (1-z_0)} \over { {(r_c -r_h)^2}(z-1)}} +{\mu^2 L^2 \over {(r_c -r_h)^2 (z-z_0)}} +(z-z_0) m^2 L^2 \bigg] } \nn \\
   &&=0\,, \nn
\end{eqnarray}
where $z_0 \equiv {r_h \over {r_h -r_c}} $ and $'$ denotes the
derivative with respect to  the variable $z$.

Defining $R(z)= z^\alpha (1-z)^\beta
(z-z_0)^\gamma \CH(z)$ and choosing parameters $\alpha$, $\beta$ and
$\gamma$ among 
\begin{eqnarray}
\{\alpha, \beta ,\gamma \}  = \bigg \{ \pm i {\w L^2 \over {(r_c -r_h)}}, ~ \pm i {\w L^2 \over {(r_c -r_h)}} ,~ \pm {\mu L \over {{{(r_c -r_h)}} \sqrt{z_0(1-z_0)}}} \bigg\} \,, \nn
\end{eqnarray}
one obtains  a differential equation for $\CH(z)$ in the form of
\begin{eqnarray} \label{heun}
 &&  \CH''(z) + \bigg[\frac{2\alpha +1}{z}
   +\frac{2 \beta +1}{z-1}+\frac{2 \gamma+1}{z-z_0}\bigg] \CH'(z) \nn \\
&& \hspace{1.1cm}  + \bigg[ \Big( (\alpha+\beta+\gamma)^2+2(\alpha+\beta+\gamma)+m^2 L^2 \Big) z  +\beta^2   -(\alpha+\gamma) (\alpha+\gamma+1)  \nonumber \\
&&\hspace{1.8cm}
 -z_0 \Big( (\alpha+\beta)^2+(\alpha+\beta)-\gamma^2  +m^2 L^2 \Big)  \bigg]
   {{\CH(z) } \over {{z (z-1)(z-z_0)}}} =0 \,.
\end{eqnarray}
This differential equation turns out to be a Heun's differential
equation~\cite{Heun} (see appendix A). We should impose the boundary condition of the ingoing modes at the   black hole horizon, $r_h$, and outgoing modes at the cosmological horizon, $r_c$. But, even though there is a local solution near $z=0$, the connection formula between Heun functions near $z=0$ and $z=1$ is not known so far.  
Therefore, by using the monodromy method we will find the asymptotic QNMs of the new type-dS black holes in the next section.

%%%%%%%%%%%%%%%%%%%%%%%%%%%%%%%%%%%%%%%%%%%%%%%%%%%%%%%%%%%%%%%%%%%%%%%%%%%%%
\subsection{Asymptotic quasinormal modes of  the new type-dS black holes}
%%%%%%%%%%%%%%%%%%%%%%%%%%%%%%%%%%%%%%%%%%%%%%%%%%%%%%%%%%%%%%%%%%%%%%%%%%%%%
%
In general, the QNMs imply damped modes with a complex frequency. To obtain the asymptotic QNMs, the monodromy method has been used as a very powerful tool in lots of literatures, for examples, see refs.~\cite{Natario:2004jd,Ghosh:2005aq}. To calculate monodromy on a closed contour, the Stokes line  along which the modes are purely oscillating  should be identified. 
Using the tortoise coordinates, we obtain the  Schr$\ddot {\rm o}$dinger-like wave equation from the eq.(\ref{wave-eq}).
Therefore, the tortoise coordinates is taken as usual such that 
 \beq \frac{dx}{dr} = \frac{L^2}{(r_c-r)(r-r_h)}\,. \eeq
By choosing the integration constant such as $x(r=0)= 0$, one obtains
\beq 
\label{ttc}
x  =  \frac{L^2}{r_c -r_h}\ln\frac{r-r_h}{r - r_c} +x_0\,,              
\eeq
where   $x_0$ is given by
\beq
 x_0  \equiv \frac{L^2}{r_c - r_h}\ln \Big( \frac{r_c}{r_h} \Big)
     = \frac{1}{2\kappa_h}\ln \Big( \frac{r_c}{r_h} \Big)\,. \nn
\eeq
By denoting $\Phi\big(x(r)\big) \equiv \sqrt{r} R(r)$, the radial
equation  of the scalar field equation can be  written as
\begin{equation}\label{eq-toto}
  \bigg[\frac{d^2}{dx^2} + \omega^2 -U(x)\bigg] \Phi(x) =0\,,
\end{equation}
where the effective potential term is given by
\begin{equation}
\label{effpoten}
\hspace{-1cm}
 U(x) = {1 \over L^2} \Big(1-\frac{r_c}{r}\Big)\Big(1-\frac{r_h}{r}\Big)
         \bigg[-\mu^2   - m^2  r^2+ \frac{1}{4 L^2}
         \Big(3r^2 - (r_c +r_h)r - r_c r_h\Big)\bigg].
\end{equation}
For the asymptotic QNMs, we consider ${\frak I}m(\omega) \gg {\frak R}e(\omega)$, with ${\frak I}m(\omega) \rightarrow \infty$, and therefore $\omega$ is approximately purely imaginary, so that $x \in i {\mathbb R}$ along the Stokes line. Near the origin, the following relation can be read from the eq.(\ref{ttc}):
\bea
r= \rho e^{i({\pi \over 2}+n \pi)} \,,
\eea
where $\rho >0$ and $n=0,1$. 
This indicates that there are two branches emanating from the origin ($n=0$ : positive branch, $n=1$ : negative branch), and  it is seen from the eq.(\ref{ttc}) that there should be a point on the real $r$-axis where a branch crosses. 
The educated guess leads to the stokes lines depicted in the figure 1.
%
% 
%
%%%%%%%%%%%%%%%%%%%%%%%%%%%%%%%%%%%%%%%%%%%%%%%%%%%%%%%%%%%%%%%
\begin{figure}  [hbtp]
\centering
\includegraphics[width=9cm,height=7cm]{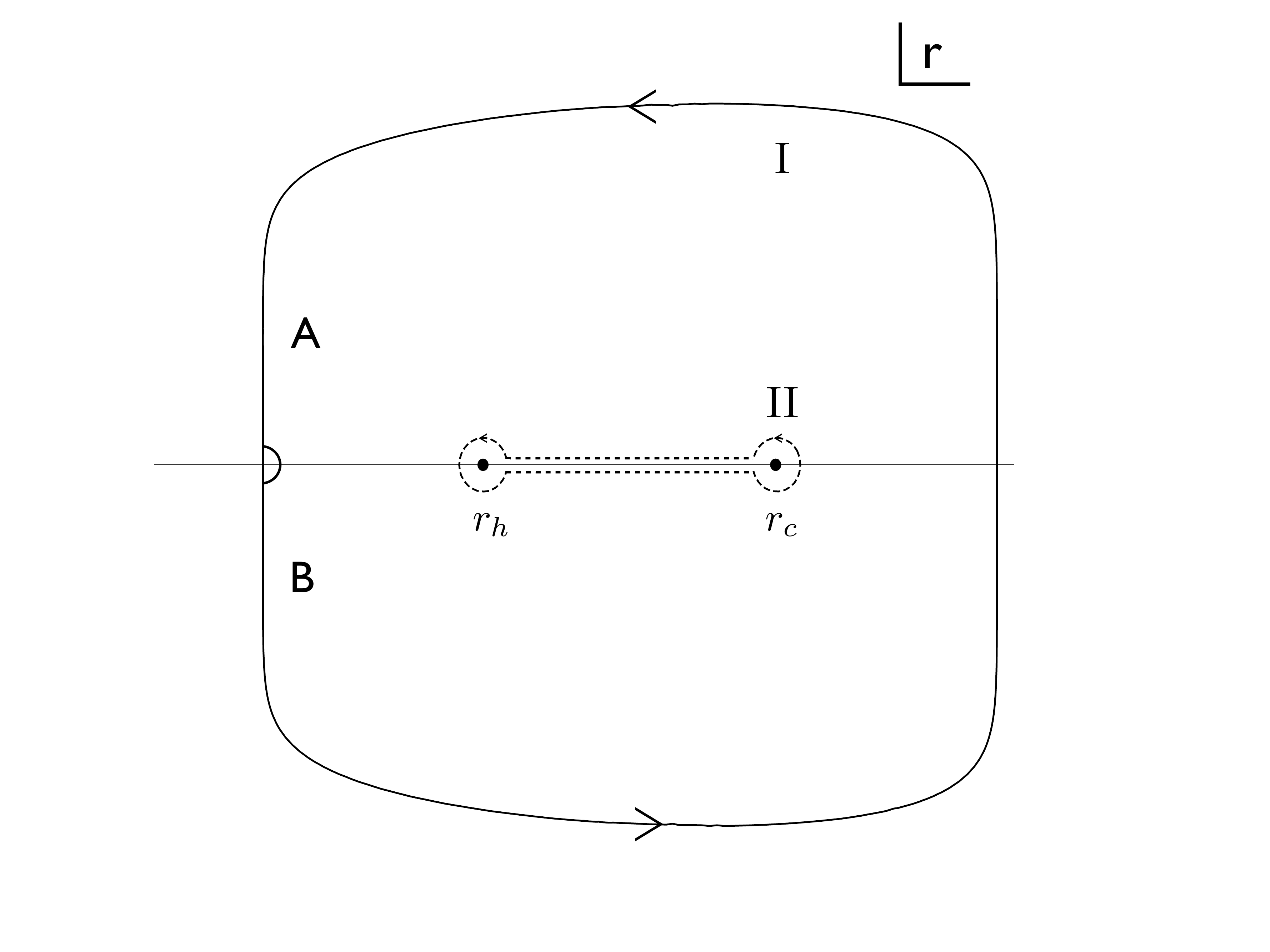}
\caption{Stokes line}
\end{figure}
%%%%%%%%%%%%%%%%%%%%%%%%%%%%%%%%%%%%%%%%%%%%%%%%%%%%%%%%%%%%%%%%
%
%   
%    
%     
%   

In the region near $r=0$, the effective potential of the  Schr$\ddot {\rm o}$dinger-like wave equation is given by 
\bea
U(x) \sim {{j^2-1} \over {4 x^2}} \,,
\eea
where $j^2 \equiv \frac{-4 \mu^2 L^2}{r_c r_h}$,
and the solution is given by
\bea
\Phi(x) \sim  A_+ \sqrt{2\pi\omega x} J_{\frac{j}{2}}(\omega x)
      + A_- \sqrt{2\pi\omega x} J_{-\frac{j}{2}}(\omega x) \,
\eea (see appendix B).
The asymptotic expansion of the above wave solution is obtained as
\begin{eqnarray}
\Phi(x) &\sim& 2 A_+ \cos(\omega x -\alpha_+)+2A_- \cos(\omega x -\alpha_-) \nn \\
&=&  (A_+ e^{-i \alpha_+} +A_- e^{-i \alpha_-} ) e^{i \omega x} +  (A_+ e^{i \alpha_+} +A_- e^{i \alpha_-} ) e^{-i \omega x} \,,
\end{eqnarray}
where $\alpha_{\pm} = {\pi}(1 \pm j)/4 $.
When one rotates from the branch containing the point A to the branch containing the point B, $x \sim - {{r L^2} \over {r_c r_h}}$ rotates through an angle of  $-\pi$.
Therefore, we have
\bea
\sqrt{2 \pi e^{-i \pi} \omega x } J_{\pm {j \over 2}} (e^{-i \pi} \omega x) = e^{-i {\pi \over 2} (1 \pm j)} \sqrt{2 \pi \omega x}  J_{\pm {j \over 2}} (\omega x) \sim 2 e^{-i 2 \alpha_{\pm}} \cos(\omega x-\alpha_{\pm}) \,. \nn
\eea
Since $e^{-i \pi} \omega x =- \omega x$, at the point B we have
\begin{eqnarray}
\Phi(x) &\sim& 2 A_+ e^{-i 2 \alpha_+} \cos(-\omega x -\alpha_+)+2A_- e^{-i 2 \alpha_-} \cos(-\omega x -\alpha_-) \nn \\
&=& (A_+ e^{-i \alpha_+} +A_- e^{-i \alpha_-} ) e^{i \omega x} +  (A_+ e^{-i 3 \alpha_+} +A_- e^{-i 3 \alpha_-} ) e^{-i \omega x}  \,.
\end{eqnarray}
With this, we can consider the contour (I) without crossing any singularities and compute monodromy corresponding to this contour.
On the one hand, we can also consider a small contour (II) equivalent to the contour (I) around singular points $r_c$ and $r_h$. 
Then the monodromy matching conditions are given by
\begin{eqnarray}
e^{ -{{\pi \omega} \over  \kappa_h} - {{\pi \omega} \over  \kappa_c}} &=&e^{ -{{\pi \omega} \over  \kappa_h} + {{\pi \omega} \over  \kappa_c}} \,,\nn \\
{ { A_+ e^{-i 3 \alpha_+} +A_- e^{-i 3 \alpha_-} } \over {A_+ e^{i \alpha_+} +A_- e^{i \alpha_-}}} e^{{{\pi \omega} \over  \kappa_h} +{{\pi \omega} \over  \kappa_c}} &=&e^{- {{\pi \omega} \over  \kappa_h}+ {{\pi \omega} \over  \kappa_c} } \,.
\end{eqnarray}
For nontrivial solutions $A_+$ and $A_-$, the following condition should be satisfied.
\bea
1-e^{{2 \pi \omega} \over {\kappa_c}} =0 \,.
\eea
Therefore, from this the QNMs of the new type-dS black holes for $0<q<1$ are given by
\bea \label{ntqnm}
\omega = i \kappa_c n =i 2\pi T_c n = i \left({{r_c-r_h} \over { 2 L^2}} \right) n \,,  \quad n=0,1,2, ...
\eea
Interestingly, it is found that the QNMs of the new type-dS black holes only depend on the Hawking temperature at the cosmological horizon, $r_c$, in contrast to the Schwarzschild-dS black hole case in higher dimensional spacetimes which has something to do with the both horizons \cite{
Natario:2004jd, Ghosh:2005aq}.

%%%%%%%%%%%%%%%%%%%%%%%%%%%%%%%%%%%%%%%%%%%%%%%%%%%%%%%%%%%%%%%%%%%%%%%%%%%%%%%%
\section{Entropy spectrum from the asymptotic quasinormal modes}
%%%%%%%%%%%%%%%%%%%%%%%%%%%%%%%%%%%%%%%%%%%%%%%%%%%%%%%%%%%%%%%%%%%%%%%%%%%%%%%%

There have been a lot of investigations about the quantizations of black holes in various methods. 
At first, Bekenstein showed that the black hole horizon area should be linearly quantized as an adiabatic invariant as follows \cite{be}:
\beq
A_n =\gamma \, n \,\hbar =\gamma \, n\, { \ell_p ^2} ~~,~~ n=0,1,2..\,,
\eeq
where $\gamma$ is an undetermined dimensionless constant, $\ell_p$ is the Planck length. %, and the $c=G=1$ unit is used. 
Later, by considering the asymptotic QNMs of a black hole, it was proposed that the transition frequency in quantum levels of black holes is relevant  to the imaginary part of the QNMs in the semi-classical limit \cite{Maggiore:2007nq}.
By identifying an action variable in a black hole system with this transition frequency, it was proposed that the following quantization condition is obtained via the Bohr-Sommerfeld quantization in the semi-classical limit \cite{Kwon:2010um}:
\bea
{\cal J}= \int { dE \over {\w_c}} = k \hbar \,,
\eea
where $E$ is the energy of a black hole system, $\w_c$ is the transition frequency defined as $\w_c \simeq |\w_I|_{n+1} -|\w_I|_{n}$ for very large $n$ ($n \gg 1$) and $k=0,1,2,3 \cdots$.
Therefore, from the above formula we will find the entropy spectra of Schwarzschild-dS and new type-dS black holes. 
First, the transition frequency of the Schwarzschild-dS is given by the eq.(\ref{sdsqnm}) as
\beq
\w_c \simeq  {{2 \sqrt{8 GM}} \over L} \,.   
\eeq
With the energy (\ref{dsen}), the quantization condition is obtained as follows:
\bea
{\cal J}= \int { dE \over \w_c} % =L \Big( 1-{1 \over { 2 m^2 L^2}} \Big) \int { dM \over {2 \sqrt{8GM}}}
=  \Big( 1-{1 \over { 2 m^2 L^2}} \Big) {{L\sqrt{8GM}} \over {8G}}= k \hbar \,,
\eea
where $k=0,1,2,3,...$.
Therefore we find the entropy (\ref{sdsen}) is quantized as
\bea
S^{c}= 4 \pi k \,.
\eea
This entropy spectrum is equally spaced and independent of the coupling parameter, $m$. Note that this is 
the same with the non-rotating BTZ black holes in the Einstein gravity and higher derivative gravities such as TMG and NMG \cite{Kwon:2013tla,Kwon:2010km,Kwon:2011ey}.

For the new type-dS black holes, since the transition frequency is given by the eq.(\ref{ntqnm}) as
\beq
\w_c \simeq  {{ r_c-r_h} \over {2 L^2}} \,,
\eeq
we have the quantization condition with the energy (\ref{nten}) as follows:
\bea
{\cal J}= \int { dE \over \w_c}=  {1 \over {4 G}} (r_c-r_h)  = k \hbar \,,
\eea
where  $k=0,1,2,3,...$.
Then the entropy (\ref{ntentro}) is quantized as
\bea
S^{c}= 2 \pi k \,.
\eea
This entropy spectrum is also equally spaced and the same with the rotating BTZ black holes in the Einstein gravity and higher derivative gravities such as TMG and NMG \cite{Kwon:2013tla,Kwon:2010km,Kwon:2011ey}.

Therefore, it is found that the entropy spectra of the Schwarzschild-dS and new type-dS black holes are equally spaced. 
In particular, it was found in lots of literatures \cite{Kwon:2011ey,Kwon:2010um,Kwon:2013tla,Kwon:2010km,Kwon:2011zza,Wei:2010yx,Chen:2010cp,LopezOrtega:2010wx,Li:2012qc,Banerjee} that the entropies of black holes in diverse gravity theories are quantized with same spacing.
Our results are also consistent with the universal property of equidistant entropy spectrum.

%%%%%%%%%%%%%%%%%%%%%%%%%%%%%%%%%%%%%%%%%%%%%%%%%%%%%%%%%%%%%%%%%%%%%
\section{Conclusion}
%%%%%%%%%%%%%%%%%%%%%%%%%%%%%%%%%%%%%%%%%%%%%%%%%%%%%%%%%%%%%%%%%%%%%

In the Einstein gravity, there was no black hole with asymptotically dS in three dimensional spacetime. In NMG with positive cosmological constant, however, such black hole solutions  can exist at the special point, $2m^2 L^2=-1$. These new type-dS black holes can be considered as the counterpart of the new type black holes with asymptotically AdS in NMG with negative cosmological constant. We have investigated some physical properties of the Schwarzschild-dS and the new type-dS black holes. By obtaining the thermodynamic quantities of these spaces, we have checked that the first thermodynamic law, $T dS =dE$ is satisfied. For this, we have calculated the energies of the Schwarzschild-dS and new type-dS black holes from the renormalized BY boundary stress tensor on the Euclidean surface at late time infinity, $\cal I^+$. 

For the Schwarzschild-dS space, it is found that  there is the upper bound in energy since the existence of the conical singularity (or black hole in higher dimensions) makes a contribution to the energy reduction from the energy of the pure dS space \cite{Balasubramanian:2001nb}. With this  interpretation, we can also consider the energy of the asymptotically dS space in higher dimensional case. For the four dimensional Schwarzschild-dS black hole, for example, the energy is given by negative value, $-m$, with black hole parameter $m$ in the metric by calculating the BY boundary stress tensor \cite{Balasubramanian:2001nb}. This can be interpreted as the subtraction of the black hole energy, $m$, from the vanishing energy of the four dimensional pure dS space \cite{Balasubramanian:2001nb}. This positive black hole energy can be obtained by using another method, so-called AD mass \cite{Abbott:1981ff,Cai:2001tv}. As the cosmological constant goes to zero, this reduces to the ADM mass in asymptotically flat spacetimes.  
For the new type-dS black boles, however, there is no upper bound in energy since the black hole parameter regime does not allow the pure dS space.

We have also obtained the QNMs of the Schwarzschild-dS and new type-dS black holes. For the Schwarzschild-dS, it is found that the scalar field equation becomes hypergeometric differential equation and the exact quasinormal modes, which is pure imaginary, was obtained. 
However, for the new type-dS black holes the scalar field equation becomes Heun's differential equation. Since the connection formula of the local solutions near singular points has not been known, we could obtain the asymptotic QNMs by using the monodromy method. 
Interestingly, the asymptotic QNMs of the new type-dS black holes, which is also pure imaginary, only depend on the Hawking temperature at the cosmological horizon, in contrast to the higher dimensional Schwarzschild-dS black holes of which algebraic equation of the asymptotic QNMs depends on the Hawking temperatures at both  black hole and cosmological horizons. By using this asymptotic QNMs, we have obtained the entropy spectrum via the Bohr-Sommerfeld quantization in the semi-classical limit. It was found that the entropy spectra of the Schwarzschild-dS and new type-dS black holes are equally spaced; $\varDelta S=4 \pi$  and $\varDelta S=2 \pi$, respectively. In these cases, we also observe the universal behavior that a black hole entropy spectrum is equally spaced.

%{\acknowledgment}
%%%%%%%%%%%%%%%%%%%%%%%%%%%%%%%%%%%%%%%%%%%%%%%%%%%%%%%%%%%%%%%%%
\section*{Acknowledgements}
%%%%%%%%%%%%%%%%%%%%%%%%%%%%%%%%%%%%%%%%%%%%%%%%%%%%%%%%%%%%%%%%%

%\ack

S.N and Y.K were supported by the National Research Foundation of
Korea(NRF) grant funded by the Korea government(MSIP) through the
Center for Quantum Spacetime (CQUeST) of Sogang University with grant
number 2005-0049409 and also supported by Basic Science
Research Program through the National Research Foundation of
Korea(NRF) funded by the Ministry of Education (No.2013R1A1A2004538).
 J.D.P was supported by a grant from 
the Kyung Hee University in 2009 (KHU-20110060) and also supported by Basic Science
Research Program through the National Research Foundation of
Korea(NRF) funded by the Ministry of Education (No.2012R1A1A2008020). 

{\it Note added} : While we are preparing this paper, the preprint  \cite{deBuyl:2013ega} 
appeared in the arXiv which deals with the thermodynamics of the dS black holes of NMG in a different way from our approach.

\newpage
\appendix
%%%%%%%%%%%%%%%%%%%%%%%%%%%%%%%%%%%%%%%%%%%%%%%%%%%%%%%%%%%%%%%%%%%%%
 %\renewcommand{\theequation}{A.\arabic{equation}}
  %\setcounter{equation}{0}
%\section{Generalized Gibbons-Hawking term}
%%%%%%%%%%%%%%%%%%%%%%%%%%%%%%%%%%%%%%%%%%%%%%%%%%%%%%%%%%%%%%%%%%%%%%

%%%%%%%%%%%%%%%%%%%%%%%%%%%%%%%%%%%%%%%%%%%%%%%%%%%%%%%%%%%%%%%%%%%%%%
\section{Heun's differential equation}
%%%%%%%%%%%%%%%%%%%%%%%%%%%%%%%%%%%%%%%%%%%%%%%%%%%%%%%%%%%%%%%%%%%%%%
 Let us recall that  the Heun's  differential
equation in the standard form is given by
\[
  \CH''(z)+\bigg[\frac{\nu }{z}+\frac{\delta}{z-1}
   +\frac{\epsilon}{z-z_0}\bigg]  \CH'(z)
   + \frac{( \lambda  \xi z -\zeta )}{z(z-1)(z-z_0)}  \CH(z) =0 \,,
\]
with the condition  $\epsilon=\lambda +\xi -\nu -\delta+1$. There exists  a local solution of this differential equation near $z=0$ which is given by a combination of local
Heun function $H(z)$'s as follows:
\bea 
   \CH(z) &=& c_1 \CH(z_0,\zeta,\lambda,\xi,\nu,\delta |z)  \\
   &&+c_2 z^{1-\nu} \CH(z_0,\zeta-(\nu-1)(\delta z_0+\lambda+\xi-\nu-\delta+1),\xi-\nu+1,\lambda-\nu+1,2-\nu,\delta |z)  \nn
\eea
Now, one
can see that the eq.(\ref{heun}) is a Heun's differential equation by
identifying
\begin{eqnarray*}
 && \nu= 2 \alpha+1 \,, ~~ \delta = 2 \beta+1 \,, ~~
    \epsilon= 2 \gamma+1 \,,
    \\
 && \zeta =  -\beta^2+(\alpha+\gamma) (\alpha+\gamma+1) +z_0 \Big( (\alpha+\beta)^2+(\alpha+\beta)-\gamma^2 +m^2 L^2  \Big) ,
    \label{lamcon1}\\
 && \{ \lambda \,,  \xi \}
    = \Big\{ \alpha+\beta+\gamma +1 \mp \sqrt{1-m^2 L^2} \,, ~\alpha+\beta+\gamma +1 \pm \sqrt{1-m^2 L^2} \Big\}
    \label{lamcon} \\
  % && \qquad \qquad {\rm or} \\
   %&&\qquad \qquad \{ \beta+\gamma+\alpha+\sqrt{1-m^2 L^2}+1  \,,~
   %\beta+\gamma+\alpha-\sqrt{1-m^2 L^2}+1   \} \,.  \label{xicon}
\end{eqnarray*}
%

%%%%%%%%%%%%%%%%%%%%%%%%%%%%%%%%%%%%%%%%%%%%%%%%%%%%%%%%%%%%%%%%%%%%%
 %\renewcommand{\theequation}{A.\arabic{equation}}
  %\setcounter{equation}{0}
\section{The wave equations near singular points  }
%%%%%%%%%%%%%%%%%%%%%%%%%%%%%%%%%%%%%%%%%%%%%%%%%%%%%%%%%%%%%%%%%%%%%%
%
Near singular points, the tortoise coordinate $x$ (\ref{ttc}) behaves like
\begin{eqnarray}
 x \simeq \left\{
              \begin{array}{ll}
                 -\frac{L^2 r}{r_c r_h}  & \qquad    r \rightarrow 0\\
                 -\infty   & \qquad    r \rightarrow r_h   \\
                 +\infty   & \qquad    r \rightarrow r_c   \\
                 x_0 + \frac{L^2}{r}  & \qquad    r \rightarrow \infty
              \end{array}
          \right.  
\end{eqnarray}
The effective potential (\ref{effpoten}) near singular points becomes
\begin{eqnarray}\label{potent}
\hspace{-0.5cm}
 U(x) \simeq
      \left\{
          \begin{array}{ll}
            \frac{r_c r_h}{L^2 r^2} \left( -\mu^2 - \frac{r_c r_h}{4 L^2} \right)
            = \frac{j^2-1}{4x^2}\,,  \qquad  j^2 \equiv \frac{-4 \mu^2 L^2}{r_c r_h}
            &  \qquad  r \rightarrow 0 ~~(x \rightarrow 0)  \\ \\
            0 & \qquad  r \rightarrow r_h  ~~(x \rightarrow -\infty)   \\  \\
             0 & \qquad  r \rightarrow r_c  ~~(x \rightarrow +\infty)   \\  \\
            \big(\frac{3}{4 L^2} - m^2 \big) {r^2 \over L^2}
            = \frac{j^2_{\infty}-1}{4(x-x_0)^2}  \,, \qquad
            j_{\infty} \equiv  2\sqrt{1-m^2 L^2}    & \qquad  r \rightarrow  \infty
            ~~(x \rightarrow x_0)
          \end{array}
      \right. 
\end{eqnarray}
Therefore, in each  singular point region, the radial function,
$\Phi(x)$, can be obtained as follows:
\begin{eqnarray}\label{sols}
\hspace{-1.5cm}
 \Phi(x) = \left\{
      \begin{array}{ll}
      A_+ \sqrt{2\pi\omega x} J_{\frac{j}{2}}(\omega x)
      + A_- \sqrt{2\pi\omega x} J_{-\frac{j}{2}}(\omega x) \,,
      &  \qquad  r \rightarrow 0 ~~(x \rightarrow 0) \\   & \\
      C_+ \sqrt{2\pi\omega (x_0-x)}
      J_{\frac{j_{\infty}}{2}}(\omega(x_0-x))   &    \\
         ~~~~~~~~~~   + C_- \sqrt{2\pi\omega (x_0-x)}
      J_{\frac{-j_{\infty}}{2}}(\omega(x_0-x)) \,,
      & \qquad  r \rightarrow  \infty ~~(x \rightarrow x_0) \\ & \\
      D_+ e^{i\omega x}  + D_- e^{-i\omega x}\,,
      & \qquad  r \rightarrow r_h  ~~(x \rightarrow -\infty) \\& \\
      F_+ e^{i\omega x}  + F_- e^{-i\omega x}\,,
      & \qquad  r \rightarrow r_c  ~~(x \rightarrow +\infty)
      \end{array}
     \right.
\end{eqnarray}
where $A_\pm$,$C_\pm$, $D_\pm$ and   $F_\pm$ are complex constants and
$J_{\pm\frac{j}{2}}(\w x)$ and
$J_{\pm\frac{j_{\infty}}{2}}(\w(x_0-x))$ represent first kind Bessel
functions.

\newpage
%\section*{References}
%%%%%%%%%%%%%%%%%%%%%%%%%%%%%%%%%%%%%%%%%%%%%%%%%%%%%%%%%%%%%%%%%%%%%

\end{document}